\documentclass[a4paper,fleqn]{cas-dc}
\usepackage{slashbox}
\usepackage{datetime}
\usepackage[numbers]{natbib}
\usepackage{bm}
\usepackage{physics}
\usepackage{siunitx}
\usepackage{amsmath}
\usepackage{nccmath}
\usepackage{graphicx}
\usepackage{amsfonts}
\usepackage{hyperref} 
\definecolor{linkcolour}{rgb}{0.,0.,0.8} 
\hypersetup{colorlinks=true,urlcolor=linkcolour,linkcolor=linkcolour,citecolor=linkcolour} 

\begin{document}
\let\WriteBookmarks\relax
\def\floatpagepagefraction{1}
\def\textpagefraction{.001}
\shorttitle{A search for neutron to mirror-neutron oscillations}
\shortauthors{\href{https://www.psi.ch/en/nedm}{nEDM collaboration at PSI}}

\title[mode = title]{
A search for neutron to mirror-neutron oscillations using the nEDM apparatus at PSI
}

\author[1]{C. Abel}

\author[2]{N.\,J. Ayres}

\author[3]{G. Ban}

\author[4]{G. Bison}

\author[5]{K. Bodek}

\author[2]{V. Bondar}

\author[6]{E. Chanel}

\author[2,4]{P.-J. Chiu}

\author[7]{C. Crawford}

\author[4]{M. Daum}

\author[8]{R.\,T. Dinani}

\author[2]{S. Emmenegger}

\author[3]{P. Flaux}

\author[9]{L. Ferraris-Bouchez}
 
\author[1]{W.\,C. Griffith}

\author[10]{Z.\,D. Gruji\'c}
\fnmark[1]

\author[2,4]{N. Hild}

\author[2,4]{K. Kirch}

\author[4]{H.-C. Koch}

\author[8]{P.\,A. Koss}

\author[11]{A. Kozela}

\author[2]{J. Krempel}

\author[4]{B. Lauss}

\author[3]{T. Lefort}

\author[9]{A. Leredde}

\author[2,4]{P. Mohanmurthy}
\cormark[1]
\fnmark[2]

\author[3]{O. Naviliat-Cuncic}

\author[2,4]{D. Pais}

\author[6]{F.\,M. Piegsa}

\author[9]{G. Pignol}

\author[2,4]{M. Rawlik}

\author[9]{D. Rebreyend}

\author[2,4]{I. Rien\"{a}cker}

\author[12]{D. Ries}

\author[9,13]{S. Roccia}

\author[5]{D. Rozpedzik}

\author[4]{P. Schmidt-Wellenburg}

\author[8]{N. Severijns}

\author[6]{J. Thorne}

\author[10]{A. Weis}

\author[8]{E. Wursten}
\fnmark[3]

\author[5]{J. Zejma}

\author[4]{G. Zsigmond}
\cormark[1]

\address[1]{Department of Physics and Astronomy, University of Sussex, Falmer, Brighton BN1 9QH, United Kingdom}
\address[2]{Institute for Particle Physics and Astrophysics, ETH Z{\"u}rich, 8093 Z{\"u}rich, Switzerland}
\address[3]{Normandie Universit\'e, ENSICAEN, UNICAEN, CNRS/IN2P3, LPC Caen, 14000 Caen, France}
\address[4]{Paul Scherrer Institut, 5232 Villigen PSI, Switzerland}
\address[5]{Marian Smoluchowski Institute of Physics, Jagiellonian University, 30-348 Krak\'ow, Poland}
\address[6]{Laboratory for High Energy Physics and Albert Einstein Center for Fundamental Physics, University of Bern, 3012 Bern, Switzerland}
\address[7]{University of Kentucky, Lexington, KY 40508, United States of America}
\address[8]{Institute for Nuclear and Radiation Physics, KU Leuven, 3001 Heverlee, Belgium}
\address[9]{Laboratoire de Physique Subatomique et de Cosmologie, Universit{\'e} Grenoble Alpes, CNRS/IN2P3, Grenoble, France}
\address[10]{University of Fribourg, 1700 Fribourg, Switzerland}
\address[11]{Henryk Niedwodnicza\'nski Institute of Nuclear Physics, 31-342 Krak\'ow, Poland}
\address[12]{Department of Chemistry - TRIGA site, Johannes Gutenberg University Mainz, 55128 Mainz, Germany}
\address[13]{Institut Laue-Langevin, CS 20156 F-38042 Grenoble Cedex 9, France}

\fntext[fn1]{Present address: Institute of Physics, University of Belgrade, 11080 Belgrade, Serbia}
\fntext[fn2]{Present address: MIT, 77 Mass. Ave., Cambridge, MA 02139, USA}
\fntext[fn3]{Present address: CERN, 1211 Gen\`eve, Switzerland}

\cortext[cor1]{Corresponding authors: geza.zsigmond@psi.ch (G. Zsigmond), \newline prajwal.thyagarthi.mohanmurthy@alumni.ethz.ch (P. Mohanmurthy)}

\begin{abstract}
It has been proposed that there could be a mirror copy of the standard model particles, restoring the parity symmetry in the weak interaction on the global level. Oscillations between a neutral standard model particle, such as the neutron, and its mirror counterpart could potentially answer various standing issues in physics today. Astrophysical studies and terrestrial experiments led by ultracold neutron storage measurements have investigated neutron to mirror-neutron oscillations and imposed constraints on the theoretical parameters. Recently, further analysis of these ultracold neutron storage experiments has yielded statistically significant anomalous signals that may be interpreted as neutron to mirror-neutron oscillations, assuming nonzero mirror magnetic fields. The neutron electric dipole moment collaboration performed a dedicated search at the Paul Scherrer Institute and found no evidence of neutron to mirror-neutron oscillations. Thereby, the following new lower limits on the oscillation time were obtained: $\tau_{nn'} > 352~$s at $B'=0$ (95\% C.L.), $\tau_{nn'} > 6~\text{s}$ for $\SI{0.4}{\micro T}<B'<\SI{25.7}{\micro T}$ (95\% C.L.), and $\tau_{nn'}/\sqrt{\cos\beta}>9~\text{s}$ for $\SI{5.0}{\micro T}<B'<\SI{25.4}{\micro T}$ (95\% C.L.), where $\beta$ is the fixed angle between the applied magnetic field and the local mirror magnetic field, which is assumed to be bound to the Earth. These new constraints are the best measured so far around $B'\sim\SI{10}{\micro T}$ and $B'\sim\SI{20}{\micro T}$.
\end{abstract}

\begin{keywords}
Properties of neutrons \sep Ultracold neutrons \sep Nuclear matter \sep Mirror matter \sep Dark matter \sep Particle symmetries
\end{keywords}

\maketitle

\section{Introduction}
Lee and Yang noted, in their landmark paper~\cite{[1]}, that parity symmetry in the weak interaction could be restored with the introduction of a parity conjugated copy of the same weakly interacting particles. It was shown by Kobzarev, Okun and Pomeranchuk~\cite{[2]} that ordinary particles would not interact with their \emph{mirror} counterparts, as they called them, via the strong, weak and electromagnetic interactions. Mirror particles would have their own interactions of the identical types \emph{i.e.} also implying the existence of mirror photons and mirror electromagnetic fields.
Foot and Volkas~\cite{Foot1991,Foot1992} detailed the aforementioned idea that by the introduction of mirror matter (hereafter denoted by SM' in analogy to SM particles), parity and time reversal symmetries could be restored in the electroweak interactions, and thus in a global sense as well. 

Several works considered that mixing of SM and SM$^{\prime}$ particles could provide answers to a number of outstanding issues in physics today.  Mirror matter could provide a viable dark matter candidate~\cite{Khlopov1989,Hodges1993,Berezhiani1996,Berezhiani2001,Berezhiani2003,Berezhiani2006}
 (for direct detection possibilities see~\cite{Foot2012,Foot2014,Addazi2015,Cerulli2017}). 
Mixings between neutrinos and mirror neutrinos~\cite{Akhmedov1992,Berezhiani1995,Silagadze1997,Berezinsky2003} due to new feeble interactions could make mirror neutrinos to natural candidates for sterile neutrino species. Furthermore, interactions of SM and SM$^{\prime}$ particles with baryon/lepton number and CP violation could open co-baryogenesis channels, thereby helping to explain the baryon dark matter fractions in the universe~\cite{Bento2001,Bento2002}. A mechanism to relax the Greisen-Zatsepin-Kuzmin (GZK) limit on the maximum energy of cosmic rays through neutron to mirror-neutron oscillations was also proposed~\cite{Berezhiani2006PLB,Berezhiani2012}. 
A comprehensive review of mirror matter physics and cosmology can be found in Refs.~\cite{Foot2014,Berezhiani2004,Berezhiani2005,Berezhiani2008,Berezhiani2018}.  

Mechanisms creating mirror magnetic fields ($B'$) on the Earth, in the solar system or Galaxy are discussed in section 4. of Ref.~\cite{[8-2]}. This suggests the possibility of $B'$-s bound to Earth of the order of \SI{100}{\micro T} which could be tested in neutron experiments.

Berezhiani and Bento~\cite{[8]} pointed out that the characteristic time for neutron to mirror-neutron ($n-n'$) oscillation, $\tau_{nn'}$, can be of the order of a few seconds, \emph{i.e.} small compared to the lifetime of the neutron.  In Ref.~\cite{[8-2]}, Berezhiani showed that, as long as neutrons and their mirror counterparts have the same mass, decay widths and gravitational potential, application of a magnetic field equal to the mirror magnetic field in the same place can induce a degeneracy between the $\ket{n}$ and $\ket{n'}$ states. This enhances the oscillation probability resonantly as described by the non-relativistic Hamiltonian:
\begin{ceqn}
\begin{align}
\mathcal{H} = \begin{pmatrix}\
-\mu_n \bm{B \cdot {\bf \sigma}} & \epsilon_{nn'} \\
\epsilon_{nn'} & -\mu_n \bm{B' \cdot {\bf \sigma}}
\end{pmatrix}, \label{eq1}
\end{align}
\end{ceqn}
where $\mu_n = -60.3~$neV/T is the magnetic moment of the neutron, $\epsilon_{nn'}=\hbar\tau^{-1}_{nn'}$ is the mass mixing term yielding a characteristic time for the $n-n'$ oscillation, $\tau_{nn'}$, and ${\bf B^{(')}}$ is the (mirror) magnetic field vector. Equation~\eqref{eq1} employs the $2\times2$ Pauli matrices, ${\bf \sigma} = (\sigma_x,\sigma_y,\sigma_z)$. The probability of $\ket{n}$ oscillating into its mirror counterpart, $\ket{n'}$, can be written as \cite{[8-2],[8-3]}: 
\begin{eqnarray}
P^{nn'}_{BB'}(t)&=&\frac{\sin^2[(\omega-\omega')t]}{2\tau_{nn'}^2(\omega-\omega')^2}+\frac{\sin^2[(\omega+\omega')t]}{2\tau_{nn'}^2(\omega+\omega')^2}\label{eq2}\\
&+&\left(\frac{\sin^2[(\omega-\omega')t]}{2\tau_{nn'}^2(\omega-\omega')^2}-\frac{\sin^2[(\omega+\omega')t]}{2\tau_{nn'}^2(\omega+\omega')^2}\right)\cos\beta\nonumber
\end{eqnarray}
where, $\omega^{(')} = | \mu_n B^{(')} | /2 = \SI{45.81}{(\micro T\cdot\text{s})^{-1}}B^{(')}$ is a convenient notation for the angular frequency in the oscillating terms above, and $t$ is the time which we know the neutrons spent in the pure normal state, $\ket{n}$. We assume a fixed angle, $\beta$, between $\bm{B}$ and $\bm{B'}$, and an approximate rotational symmetry around the Earth's axis for the mirror magnetic field, subject to experimental testing. 

Neutron to mirror-neutron oscillation would manifest itself as an additional loss channel in ultracold neutron (UCN) storage experiments~\cite{[3-1]}, since if a UCN oscillates into its mirror counterpart, it would escape the storage chamber. Far away from the resonance, when for UCNs $|\omega-\omega'|t\gg1$, Eq.~\eqref{eq2} can be averaged over time and reduced to~\cite{[8-2]}:
\begin{ceqn}
\begin{align}
P^{nn'}_{BB'} = P^{nn'}_{0 B'}\frac{1+\eta^2+2\eta\cos\beta}{\left(1-\eta^2\right)^2}, \label{eq3}
\end{align}
\end{ceqn}
where $\eta=\omega/\omega'$, and 
\begin{ceqn}
\begin{align}
P^{nn'}_{0 B'} = 1/(2\tau_{nn'}^2\omega'^2)  \label{eqPforBzero}
\end{align}
\end{ceqn}
is the $n-n'$ oscillation probability in the absence of a magnetic field ($B=0$) valid for $\omega' t \gg 1$. The time $t$ is reset to zero at each wall reflection since a successful reflection confirms the neutron being a SM particle. Using the mean time between two consecutive wall-collisions $\left<t_f\right>$, the average number of free flight segments during a storage time $t_s$ can be approximated as $m_s = t_s /\left<t_f\right>$. For $P^{nn'}_{BB'}$ we consider the average over the free flight time $t_f$. The attenuation in the number of UCNs due to this loss channel is then $\exp(-m_s P^{nn'}_{BB'})$. Close to the resonance, Eq.~\eqref{eq3} has to be complemented as explained in detail in Ref.~\cite{[8-3]} to cancel out the singularity at $\omega=\omega'$.

Berezhiani et al.~\cite{[8-2]} pointed out that in order to set constraints on $\tau_{nn'}$ as a function of the mirror magnetic field it is convenient to work with the observables `ratio' ($E_B$) and `asymmetry' ($A_B$), respectively, defined as:
\begin{eqnarray}
\hspace{-8mm}E_B^{\left(t_s\right)}+1=\frac{2 n_0^{\left(t_s\right)}}{n_{B}^{\left(t_s\right)}+n_{-B}^{\left(t_s\right)}}&=\frac{2e^{-\left(m_s P^{nn'}_{0 B'}\right)}}{e^{-\left(m_s P^{nn'}_{BB'}\right)}+e^{-\left(m_s P^{nn'}_{-BB'}\right)}},\label{eq4}\\
\hspace{-8mm}A_B^{\left(t_s\right)}=\frac{n_{B}^{\left(t_s\right)}-n_{-B}^{\left(t_s\right)}}{n_{B}^{\left(t_s\right)}+n_{-B}^{\left(t_s\right)}}&=\frac{e^{-\left(m_s P^{nn'}_{BB'}\right)}-e^{-\left(m_s P^{nn'}_{-BB'}\right)}}{e^{-\left(m_s P^{nn'}_{BB'}\right)}+e^{-\left(m_s P^{nn'}_{-BB'}\right)}},\label{eq5}
\end{eqnarray}
where the $n_{\{0,B,-B\}}^{(t_s)}$ are the number of neutrons counted after storage for time $t_s$. The indices $B$ and $-B$ in the above equations refer to the direction of the applied magnetic field along the vertical axis at the location of the UCN storage chamber. The attenuation in UCN counts due to losses at wall collisions and $\beta$-decay, and  the detection efficiency are independent from the applied field $B$ and thus will cancel out from the count ratios.  

When we assume the mirror magnetic field to be zero ($B'=0$), the relationships between the $n-n'$ oscillation time, $\tau^{(B'=0)}_{nn'}$, and the ratio observable in Eq.~\eqref{eq4} becomes independent of the applied magnetic field. Considering the limits $\omega\left<t_f\right>\ll1$ (no field applied) and $\omega \left<t_f\right>\gg1$  (field applied) with $P^{nn'}_{B B'}$ $\ll1$, as in Refs.~\cite{[3-1],[9-1]}, yields: 
\begin{ceqn}
\begin{align}
&\tau^2_{nn'} \overset{\tiny{B'=0}}{\simeq} \underbrace{-t_s\frac{\left<t^2_f\right>}{\left<t_f\right>}\frac{1}{E_B}}_{-1/\Delta_0}. \label{eq4-0}
\end{align}
\end{ceqn}
Since probability and $\tau^2_{nn'}$ (see Eq.~\eqref{eqPforBzero}) are positive quantities, $\Delta_0$ is only physical for negative values (\emph{e.g.} in the limit of $B\approx 0,~B'\approx 0$, $E_{B}^{\left(t_s\right)}\approx -m_s P^{nn'}_{BB'}$). The rightmost terms in  Eqs.~\eqref{eq4}-\eqref{eq5} were defined in the context of a disappearance experiment, thus the number of SM neutrons can only decrease.

Including the case when the mirror magnetic field is non-zero, the ratio and asymmetry observables in Eqs.~\eqref{eq4} and \eqref{eq5}, respectively, are linked to the $n-n'$ oscillation time through Eq.~\eqref{eq3} as follows \cite{[8-2]}:
\begin{eqnarray}
\tau^2_{nn'} \overset{\tiny{B'\ne0}}{\simeq}& \underbrace{\frac{t_s}{\left<t_f\right>}\frac{1}{E_B}}_{1/\Delta_B}\cdot\underbrace{\frac{\eta^2\left(3-\eta^2\right)}{2\omega'^2\left(1-\eta^2\right)^2}}_{f_{E_B}(\eta)},  \label{eq4-2}\\
\frac{\tau^2_{nn'}}{\cos\beta} \overset{\tiny{B'\ne0}}{\simeq}& \underbrace{-\frac{t_s}{\left<t_f\right>}\frac{1}{A_B}}_{-1/D_B}\cdot\underbrace{\frac{\eta^3}{\omega^2\left(1-\eta^2\right)^2}}_{f_{A_B}(\eta)}, \label{eq5-2}
\end{eqnarray}
where $f_{\{E_B, A_B\}}(\eta)$ are the scaling functions. The conditions $P^{nn'}_{B B'}$ $\ll1$ and $\omega'\left<t_f\right>$ $\gg 1$ have to be fulfilled. $\Delta_0$, $\Delta_B$ and $D_B$ will be used and discussed in subsection~\ref{subsec:3.2}. The null-hypothesis is that there are no $n-n'$ oscillations, and consequently the measured value of $E_B$ and $A_B$, in Eqs.~\eqref{eq4} and~\eqref{eq5}, respectively, would be consistent with zero. Deviations from the null-hypothesis are referred to as signals.

The first series of experiments with UCNs used the ratio observable under the assumption of $B'=0$. They set the constraints of $\tau_{nn'} > 103~$s (95\% C.L.)~\cite{[9-1]} and later $\tau_{nn'} > 414~$s (90\% C.L.)~\cite{[9-2]}. Reference~\cite{[9-2]} has since updated their constraint to $\tau_{nn'} > 448~$s (90\% C.L.) \cite{[9-2-1]}. Reference~\cite{[9-3]} relaxed the conditions to $B'\ne0$, while still using the ratio observable, and set a constraint of $\tau_{nn'} > 12~\text{s}$ for $\SI{0.4}{\micro T}<B'<\SI{12.5}{\micro T}$ (95\% C.L.). In Ref.~\cite{[8-3]}, Berezhiani et al. further analyzed the above experiments and indicated statistically significant signal-like anomalies for $n-n'$ oscillation in the asymmetry observable when $B'\ne0$. The experiment presented here was designed to check the potential signals in Ref.~\cite{[8-3]}, and provide sufficient sensitivity to exclude them if not real. A recent update by Berezhiani et al. \cite{[8-4],Biondi2018} shows a persistence of the anomalous signals. Reference~\cite{[8-4]} also sets constraints of $\tau_{nn'}>17~\text{s}$ for $\SI{8}{\micro T}<B'<\SI{17}{\micro T}$ (at 95\% C.L.) and $\tau_{nn'}/\sqrt{\cos\beta}>27~\text{s}$ for $\SI{6}{\micro T}<B'<\SI{25}{\micro T}$ (at 95\% C.L.). The three statistically significant signals identified in the asymmetry (unfortunately deviating from those in Ref.~\cite{[8-3]}) are: a $3\sigma$ signal from the data in Ref. \cite{[9-1]}, a $5.2\sigma$ signal from data in Refs.~\cite{[9-2],[9-2-1]}, and a $2.5\sigma$ signal from the B2 data series in Ref.~\cite{[8-4]}.
Testing the above anomalies in the asymmetry observable of $n-n'$ oscillation was the primary motivation for this measurement at the Paul Scherrer Institute (PSI) by the neutron electric dipole moment (nEDM) collaboration. 

\section{Experiment setup and data collection}

For this experiment, the PSI collaboration made use of its repurposed nEDM apparatus described in Refs.~\cite{[10-2],[7-1-1-1],[Abel2020]} hosted at the PSI ultracold neutron source \cite{NeutronOptics2020}. A UCN guide switch directed the neutrons coming from the beamport to a $21$~liter cylindrical storage chamber. The storage chamber was made of a polystyrene insulator ring coated with deuterated polystyrene, sandwiched between two aluminum plates (the electrodes for the nEDM search) coated with diamond-like carbon \cite{[7-3-1],[7-3-2],[7-3-3]}. The storage chamber was enclosed in a vacuum tank on which a coil system was wound that generated the vertical magnetic field, $B$ (called $B_0$ in the nEDM experiment). It was surrounded by a four-layer $\mu$-metal shield which was housed inside an active magnetic field compensation system~\cite{[7-2-2]}. In this $n-n'$ oscillation search no electric field was used. The storage chamber was connected via the switch to a neutron detection system \cite{[7-1-1],[7-1-2]}. 

In this experiment we used unpolarized neutrons in order to maximize statistics. 
Data was collected in a series of runs and each run consisted of many cycles. The neutron storage time, $t^*_s$, during each cycle was fixed per run, but the magnetic field was changed from cycle to cycle in a specific pattern.
In the beginning of a cycle, the UCNs from the source were allowed to fill the storage chamber after passing through the appropriately configured switch. The UCN shutter at the bottom of the storage chamber was then shut. After a period of storage, the shutter of the storage chamber was opened and the neutrons were counted. We will refer to this part of the cycle as the emptying phase. 

In order to compensate for fluctuations of the UCN source output \cite{[11-1]} the detector counts at the end of a cycle had to be normalized using a monitor. The neutrons still emerging from the source during the storage phase were directly guided to the UCN detectors, serving as monitor counts. The monitor counts were of the order of a million; the emptying counts, after the storage, was of the order of a few tens of thousands. Thus, the uncertainty on the ratio of emptying and monitor counts is mostly dependent on the uncertainty coming from the emptying counts. Special care was taken to demonstrate that this ratio was stable enough for the $n-n'$ oscillation search as explained in Ref.~\cite{[10]}. Henceforth, we will denote the emptying counts corrected using the monitor counts as $n_{\{0,B,-B\}}^{(t_s)}$.

The data was taken with storage times, $t^*_s$, set to $180~$s and $380~$s.  The selected longer storage time is the result of an optimization for the best sensitivity to $n-n'$ oscillation \cite{[10]}, while the shorter one allowed for a direct comparison to previous measurements. In order to account for the total time the neutrons spent in the magnetic field region, we also need to consider the average time of filling and emptying the chamber. During the filling of the chamber, the UCN density builds up until it reaches equilibrium. This is characterized by an exponential time constant. The chamber is filled and emptied through the same opening and same vertical guide. Consequently, for the energy spectrum of the UCNs detected at the end of storage, the filling time constant is approximately equal to the emptying time constant. We added twice the emptying time constant of the UCNs to the storage time set in the control system: $t_s = t^*_s + 2\tau_{\text{emp}}(t^*_s)$, where $\tau_{\text{emp}}$ is the filling (or emptying) time constant. 

The magnetic field applied was calibrated using the $^{199}$Hg co-magnetometer~\cite{HgComag} of the nEDM apparatus and a nanoampere meter to measure the current supplied to the $B$ coil. Along with the $B=0$ reference case, magnetic fields of 
$\SI[separate-uncertainty = true]{10.20(2)}{\micro T}$ and $ \SI[separate-uncertainty = true]{20.39(4)}{\micro T}$ 
were used  in these measurements, optimal to address the aforementioned anomalous signals of Ref.~\cite{[8-3]}. The errors given here are larger than the inhomogeneity of the field. The requirement for precision on the magnetic field is elaborated on in Ref.~\cite{[10]}. Patterns of $16$ settings of the magnetic field, [0, B, 0, -B, 0, -B, 0, B, 0, -B, 0, B, 0, B, 0, -B], were applied by changing the magnetic field after every \emph{four} cycles. Such patterns can compensate for drifts in the magnetic field \cite{[8-1]}. One full pattern consists of $64$ cycles. We collected over $8000$ cycles of data. 

\section{Data analysis and results}

Apart from the data collected in the experiment, the analysis needs the distribution of the flight time between consecutive collisions, $t_f$, as an input. This input was provided by MC simulations fitted to measured data. Further, the data analysis focused on the two observables, the ratio and the asymmetry. The null result was interpreted by setting constraints on the $n-n'$ oscillation parameters.

\subsection{Calculation of the free flight time distributions}
\label{subsec:3.1}

\sloppy We remind that Eqs.~\eqref{eq4-0},~\eqref{eq4-2}, and~\eqref{eq5-2} use the mean time, $\left<t_f\right>^{(t_s)}$ between consecutive wall collisions. Below we summarize the steps of our method. For calculation details we refer the reader to section 3.6 in Ref.~\cite{[14]}. 

To obtain $\left<t_f\right>^{(t_s)}$ for each time of storage, $t_s$, the free flight time of UCNs had to be averaged first over the path through the chamber for each energy bin separately, and then over a given energy spectrum. The path history of UCNs in a storage chamber yielded a broad $t_f$ distribution. Neutrons bouncing at the corners of the storage chamber, or slower neutrons bouncing due to gravity along the bottom surface of the chamber, will contribute to small values of $t_f$. Neutrons traversing the longest paths in the storage chamber will contribute to larger values of $t_f$ (depending also on the magnitude of the velocity). 
While the geometry of the storage chamber determines the path length distribution as a function of energy very well, the uncertainty on $\left<t_f\right>^{(t_s)}$ is dominated by the uncertainty of the less well-known energy spectrum.

The energy spectrum and the associated uncertainties were extracted using an analytical model for the storage curve as detailed in section 3.1.3 of Ref.~\cite{[10-3]} along with simulation tests. This model involves the energy dependent bounce rate $\nu(E)$ and the loss probability per bounce $\mu(E)$~\cite{Golub1991} (their product giving the loss rate) via the decay function:
\begin{ceqn}
\begin{align}
n(E,t_s)=n(E,0)\exp(-t_s\nu(E)\mu(E)),\label{eqDecay}
\end{align}
\end{ceqn}
where $E = E_{b} - m_n g h$ denotes the kinetic energy at the average height of collision, $h$, and at the bottom of the chamber $E(h=0)=E_b$. The energy spectrum at the bottom of the chamber and at the beginning of the storage phase ($t_s=0$), was parameterized with a peak function of the form:
\begin{ceqn}
\begin{align}
P(E_{b})=P_0 \frac{E_{b}^\rho}{1+\exp(\frac{E_{b} - E_p}{w})},\label{eqSpectr}
\end{align}
\end{ceqn}
where $P_0$ is a scaling constant, $\rho$ is the exponent of the leading edge of the distribution, $E_p$ is an upper cut-off value for the energy, and $w$ is a smearing parameter for the cut-off. A similar sigmoidal definition was used in Ref.~\cite{UCNSE}.

Equation~\eqref{eqDecay} was integrated with the spectral weighting, Eq.~\eqref{eqSpectr}, using the above definition $E_{b} = E + m_n g h$:
\begin{ceqn}
\begin{align}
n(t_s) = \int n(E_{b}, t_s) P(E_{b})dE_{b}.\label{eqDecayIntegrated}
\end{align}
\end{ceqn}
We used this function to fit the storage curve, $n_{meas}(t_s)$, measured for this purpose at $15$ different storage times~\cite{[14]}. The analytical model distinguished between the average loss rates, $\nu(E)\mu(E)$ at the top, bottom, and side surfaces, adding these together. Concerning the side wall, Eq.~(5) in Ref.~\cite{Harris2014} for the average height of UCNs in a cylindrical chamber was employed. The fit to the measured storage curve was performed by randomly sampling the parameters $\{P_0,\rho,E_p,w\}$, and the wall loss parameter $\eta'=W/V$, which is the ratio of the imaginary and real parts of the optical potential of the coating material~\cite{Golub1991}. The Fisher statistical test~\cite{PDG2018} was used to obtain the confidence regions in the parameter space. 

For every set of $\{P_0,\rho,E_p,w\}$, a center of mass offset of UCNs \emph{w.r.t.} the center of the chamber, $\left<z\right>$, was calculated~\cite{[14]}. A further constraint on the $\{P_0,\rho,E_p,w\}$ parameters was imposed by using the measurement of $\left<z\right>$ in the nEDM experiment~\cite{[Abel2020]}. The nEDM search requires polarized neutrons, whereas this $n-n'$ oscillation search used unpolarized neutrons. The center of mass offset was simulated with both polarized and unpolarized neutrons from the beamline. The difference was within the error of the calculations.

\begin{figure}
\centering
	\includegraphics[width=8cm]{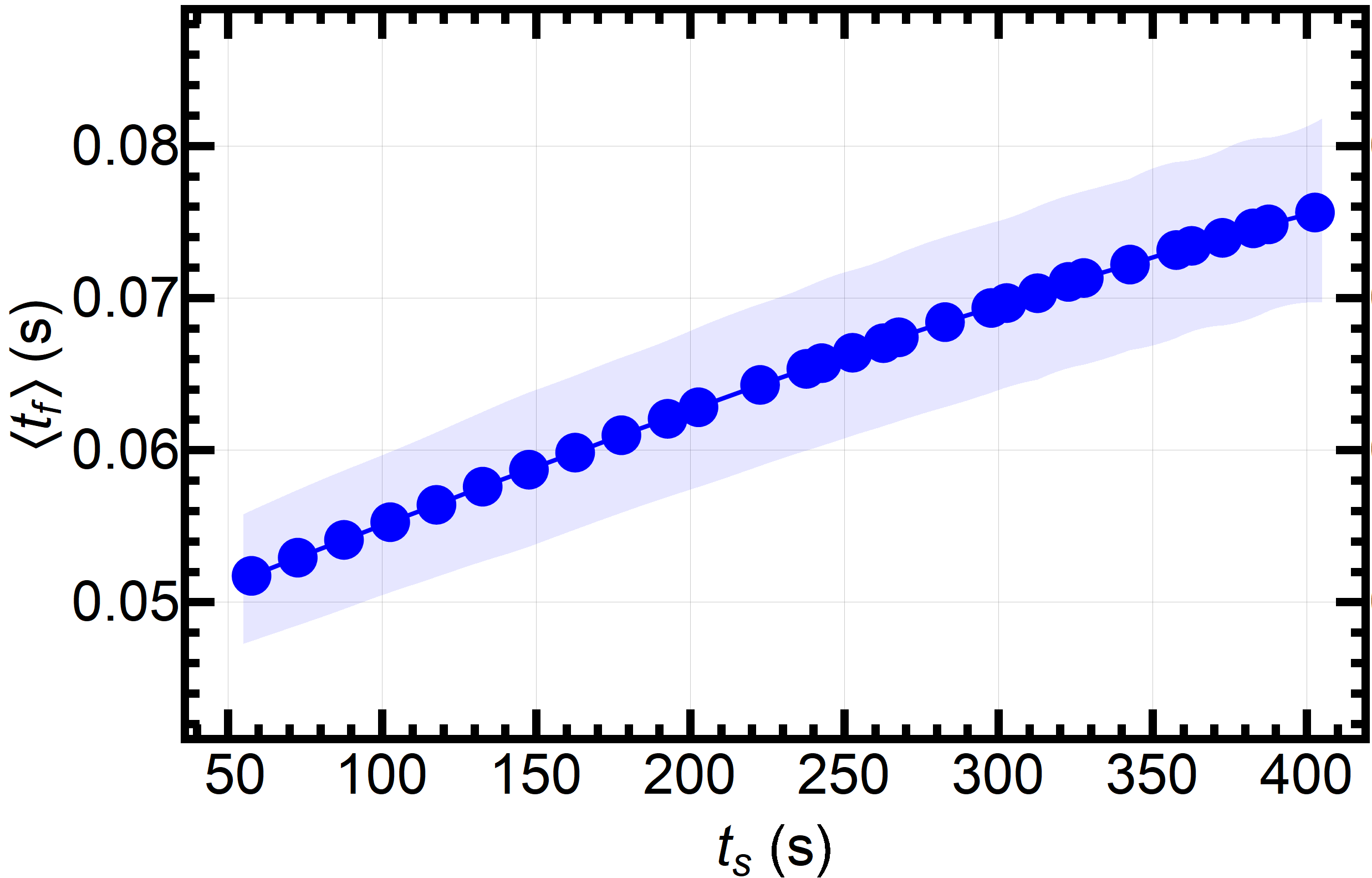}
\caption[]{Simulated dependence of $\left<t_f\right>^{(t_s)}$  \emph{w.r.t.} the storage time. The data points represent the central value of the $\left<t_f\right>^{(t_s)}$ distribution and the shaded region shows the $95$\% C.L. contours of the width of the $\left<t_f\right>^{(t_s)}$ distribution.}
\label{fig1}
\vspace{-5mm}
\end{figure}

The energy spectra associated with each set of parameters $\{P_0,\rho,E_p,w\}$ were next translated to distributions of $t_f^{(t_s)}$ by the means of ray-tracing using the MCUCN code~\cite{[10-3]}. The profiles turned out to be normal distributions. We noticed that the central values of the $\left<t_f\right>^{(t_s)}$ distributions and the associated uncertainties vary appreciably with storage time, as visible in Figure~\ref{fig1}.  This was taken into account in the analysis. The largest contributor to the width of the $\left<t_f\right>^{(t_s)}$ distribution is the uncertainty on the energy spectrum parameters. The uncertainty contribution from path averaging is much smaller, since, during the given storage times, the UCNs can bounce off the walls diffusely, a large number of times, thus achieving mechanical equilibrium. Its uncertainty is only limited by the statistical accuracy of the MC simulations.

While the $n-n'$ oscillation time in non-zero mirror magnetic fields, from Eqs.~\eqref{eq4-2}-\eqref{eq5-2} only requires $\left<t_f\right>^{(t_s)}$, in zero mirror magnetic field, Eq.~\eqref{eq4-0} requires $(\left<t_f^2\right>/\left<t_f\right>)^{(t_s)}$, and the associated uncertainty. These were calculated in a similar way by MCUCN simulations. 

As a byproduct of the energy spectrum calculations, we also obtained a constraint on the wall loss parameter of the precession chamber in the nEDM experiment. This value is effectively averaged (in proportion to the area) over the insulator ring and the electrode surfaces: $\eta' = (2.5\pm0.3)\times10^{-4}$.

\subsection{Constraints on the ratio and asymmetry observables}
\label{subsec:3.2}

Each run is associated with a storage time, $t_s$, and a maximum magnetic field, $B$, that was applied in the aforementioned pattern. Within each run the emptying counts corrected by the monitor counts cycle by cycle, $n^{(t_s)}_{\{B,0,-B\}}$, were grouped according to the three field configurations of $\{B, 0, -B\}$. Within each group the mean values and the standard errors on the mean were calculated. From these, the values $\left<E_B^{(t_s)}\right>$ and $\left<A_B^{(t_s)}\right>$ were obtained using Eqs.~\eqref{eq4} and \eqref{eq5}. The errors on the mean values, $\left<n^{(t_s)}_{\{B,0,-B\}}\right>$, were propagated to obtain the errors on $\left<E_B^{(t_s)}\right>$ and $\left<A_B^{(t_s)}\right>$. 

\begin{figure}
\raggedleft
	\includegraphics[width=83mm]{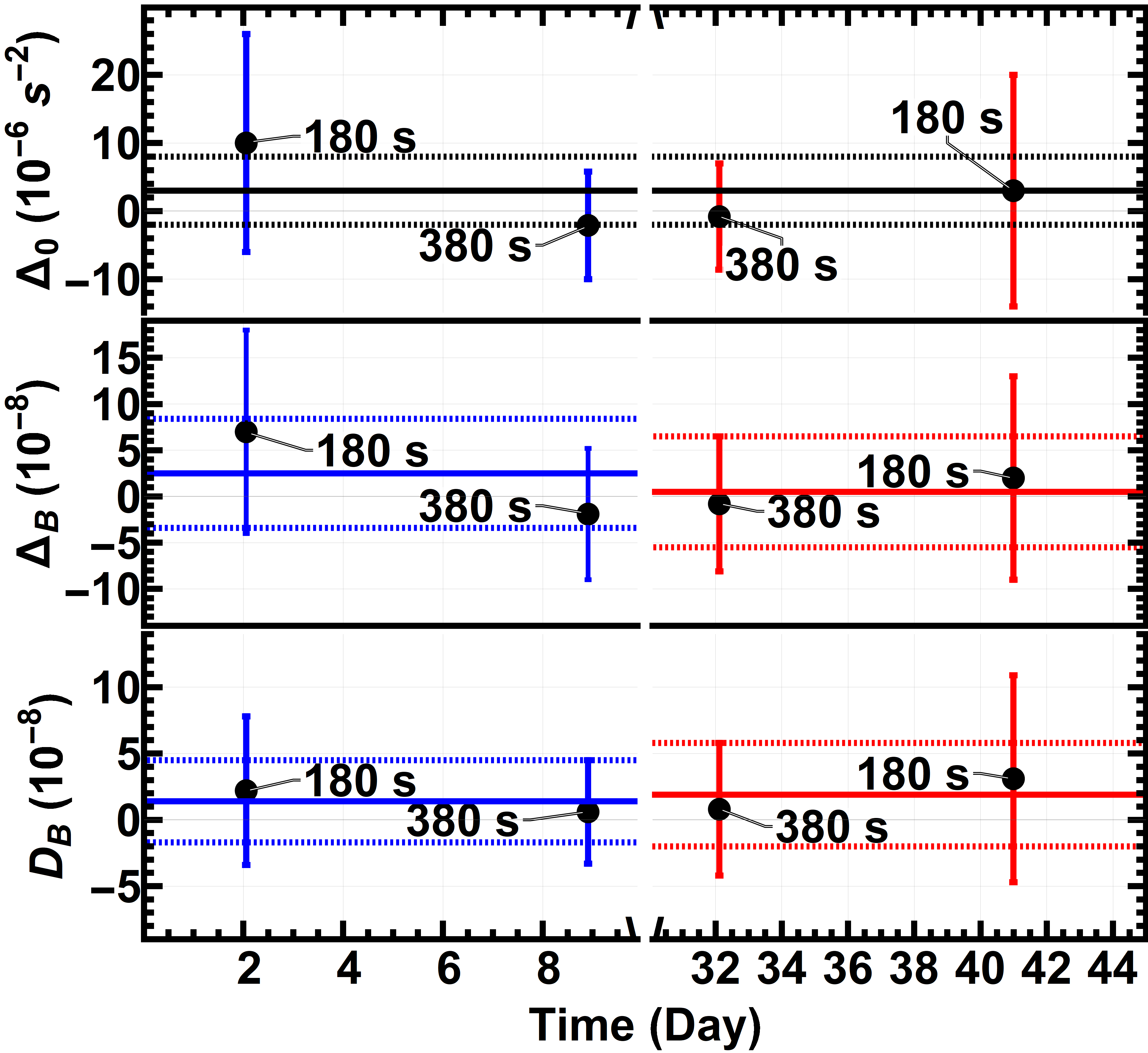}
	\caption[]{Values of $\Delta_0$ (Top), $\Delta_B$ (Center), and $D_B$ (Bottom), from Eqs.~\eqref{eq4-0}, \eqref{eq4-2}, and \eqref{eq5-2}, respectively, plotted for each run as a function of the mean time at which the data for the run was collected. The data points associated with blue error bars show those runs involving a magnetic field of $B\sim\SI{10}{\micro T}$, while the data points associated with red error bars show the runs involving a magnetic field of $B\sim\SI{20}{\micro T}$. The solid lines of the same color represent the weighted mean of the data points, and the dashed lines represent the standard errors, as listed in Eqs.~\eqref{eq6-1}-\eqref{eq6-3-2}.}
\label{fig3}
\vspace{-5mm}
\end{figure}

The terms, $\Delta_0$, $\Delta_B$, and $D_B$, in Eqs.~\eqref{eq4-0}, \eqref{eq4-2}, and \eqref{eq5-2}, respectively, allowed us to combine the various runs as in Refs.~\cite{[8-2],[8-3],[8-4]}, each with corresponding values of $t_s$ and $\left<t_f\right>^{(t_s)}$, which are shown for each run in Figure~\ref{fig3}~(Top), (Center), and (Bottom), respectively. The weighted averages and the corresponding errors for the various settings are: 

\begin{eqnarray}
\hspace{-7mm}\left<\Delta_0\right>&=&(3.0\pm5.0)\times10^{-6}~\text{s}^{-2},\label{eq6-1}\\
\hspace{-7mm}\Bigg\langle\underbrace{\left<E_{B\sim\SI{10}{\micro T}}\right>\frac{\left<t_f\right>^{(t_s)}}{t_s}}_{\Delta_{B\sim\SI{10}{\micro T}}}\Bigg\rangle&=&(2.5\pm5.9)\times10^{-8},\label{eq6-2-1}\\
\hspace{-7mm}\Bigg\langle\underbrace{\left<E_{B\sim\SI{20}{\micro T}}\right>\frac{\left<t_f\right>^{(t_s)}}{t_s}}_{\Delta_{B\sim\SI{20}{\micro T}}}\Bigg\rangle&=&(0.5\pm6.0)\times10^{-8},\label{eq6-2-2}\\
\hspace{-7mm}\Bigg\langle\underbrace{\left<A_{B\sim\SI{10}{\micro T}}\right>\frac{\left<t_f\right>^{(t_s)}}{t_s}}_{D_{B\sim\SI{10}{\micro T}}}\Bigg\rangle&=&(1.4\pm3.1)\times10^{-8},\label{eq6-3-1}\\
\hspace{-7mm}\Bigg\langle\underbrace{\left<A_{B\sim\SI{20}{\micro T}}\right>\frac{\left<t_f\right>^{(t_s)}}{t_s}}_{D_{B\sim\SI{20}{\micro T}}}\Bigg\rangle&=&(1.9\pm3.9)\times10^{-8}.\label{eq6-3-2}
\end{eqnarray}
The uncertainty associated with the values of $\left<\Delta_0\right>$, $\left<\Delta_B\right>$, and $\left<D_B\right>$ in Figure~\ref{fig3}, comes from propagating the uncertainty on the values of $\left<E_B^{(t_s)}\right>$, $\left<A_B^{(t_s)}\right>$, $t_s$ and $\left<t_f\right>^{(t_s)}$, according to Eqs.~\eqref{eq4-0}, \eqref{eq4-2}, and \eqref{eq5-2}. We emphasize here that in the calculation of the distribution parameters of $\Delta_0$, $\Delta_B$, and $D_B$ we used both positive and negative values, contrary to subsection~\ref{subsec:3.3} where these quantities are sampled either in negative or positive intervals, wherever the oscillation probability is positive.

In order to give an estimate on the uncertainty contributions to $\left<\Delta_0\right>$, $\left<\Delta_B\right>$, and $\left<D_B\right>$ separately from the emptying counts, monitor counts, $\left<t_f\right>^{(t_s)}$, and $t_s$ (via $\tau_{\text{emp}}$), we calculated  the error propagation from the definitions in Eqs.~\eqref{eq4-0}, \eqref{eq4-2} and \eqref{eq5-2}. 
The different uncertainty contributions are compared in Table~\ref{table:ContribUncert}.
\begin{table}
\centering
\begin{tabular}{|c|c|c|c|c|}
\hline
\backslashbox{Errors for}{From} & $N_{emp}$  &  $N_{mon}$  & $\left<t_f\right>$ & $t_s$ \\
\hline
$\left<\Delta_0\right>$ 	($10^{-6}s^{-2}$)     &  4.74 &	1.41 &	0.06 &	0.002  \\
$\left<\Delta_{B\sim\SI{10}{\micro T}}\right>$ 	($10^{-8}$)  &  5.51 &	1.54	& 0.07 & 0.002   \\
$\left<\Delta_{B\sim\SI{20}{\micro T}}\right>$ 	($10^{-8}$) &  5.80	 & 1.80	& 0.03 &	0.002  \\
$\left<D_{B\sim\SI{10}{\micro T}}\right>$ 	($10^{-8}$)   & 2.92	 &  0.85	& 0.02	& 0.002   \\
$\left<D_{B\sim\SI{20}{\micro T}}\right>$ 	($10^{-8}$)  &  3.76	 &  1.13	& 0.03 & 0.002\\
\hline
\end{tabular}
\caption[]{Uncertainty contributions to $\left<\Delta_0\right>$, $\left<\Delta_B\right>$, and $\left<D_B\right>$ separately from emptying counts ($N_{emp}$), monitor counts ($N_{mon}$), mean free flight time ($\left<t_f\right>$, including also $\left<t_f^2\right>$), and effective storage time ($t_s$).}
\label{table:ContribUncert}
\end{table}

We did not observe any statistically significant deviations of $\left<E_B^{(t_s)}\right>$ or $\left<A_B^{(t_s)}\right>$ from zero, and consequently the weighted means in Eqs.~\eqref{eq6-1}-\eqref{eq6-3-2} are consistent with zero. Therefore, we only present constraints on the $n-n'$ oscillation time parameter $\tau_{nn'}$.

\subsection{Constraints on the $n-n'$ oscillation time and mirror magnetic field}
\label{subsec:3.3}

By applying the constraints in Eqs.~\eqref{eq6-1}-\eqref{eq6-3-2}, we can construct exclusion diagrams in the parameter space of $n-n'$ oscillations. From Eq.~\eqref{eq4-0} we see that the $n-n'$ oscillation time under the assumption of $B'=0$ is given by the function $\tau_{nn'} = 1/\sqrt{-\left<\Delta_0\right>}$. Therefore, we numerically sampled $\Delta_0$ in the negative range of the normal distribution, to avoid imaginary numbers and negative probability, according to the parameters in Eq.~\eqref{eq6-1}, and obtained the following constraint:
\begin{ceqn}
\begin{align}
\tau^{B'=0}_{nn'} > 352~\text{s (95\% C.L.)}. \label{eq10}
\end{align}
\end{ceqn}

In case of the ratio observable, Eq.~\eqref{eq4-2}, since the sign of the function $f_{E_B}(\eta)$ changes at $B'\sqrt{3}=B$, we subsequently extracted the lower limit of $\tau_{nn'}^{B'\ne0,E_B}/\sqrt{\left|f_{E_B}(\eta)\right|}=1/\sqrt{\left<\Delta_B\right>}$ using both the distributions of $\left<\Delta_B\right>$ and $-\left<\Delta_B\right>$, in their appropriate ranges, to avoid imaginary numbers for the oscillation time, $\tau_{nn'}^{B'\ne0,E_B}$. Similar to the case where we assumed $B'=0$, the weighted averages in Eqs.~\eqref{eq6-2-1}-\eqref{eq6-2-2} were numerically sampled to obtain the following constraints, at $95\%$ C.L.:
\begin{eqnarray}
\frac{\tau_{nn'}^{B'\ne0,E_B}}{\sqrt{\left|f_{E_B}(\eta)\right|}}&>3145~(B\sim\SI{10}{\micro T},~B'\sqrt{3}< B) \label{eq5-7-3-1-6},\\
&>2948~(B\sim\SI{20}{\micro T},~B'\sqrt{3}< B) \label{eq5-7-3-1-7},\\
&>2954~(B\sim\SI{10}{\micro T},~B'\sqrt{3}> B)\label{eq5-7-3-1-8},\\
&>2914~(B\sim\SI{20}{\micro T},~B'\sqrt{3}> B) \label{eq5-7-3-1-9}.
\end{eqnarray}

\begin{figure*}
\centering
\includegraphics[width=\textwidth]{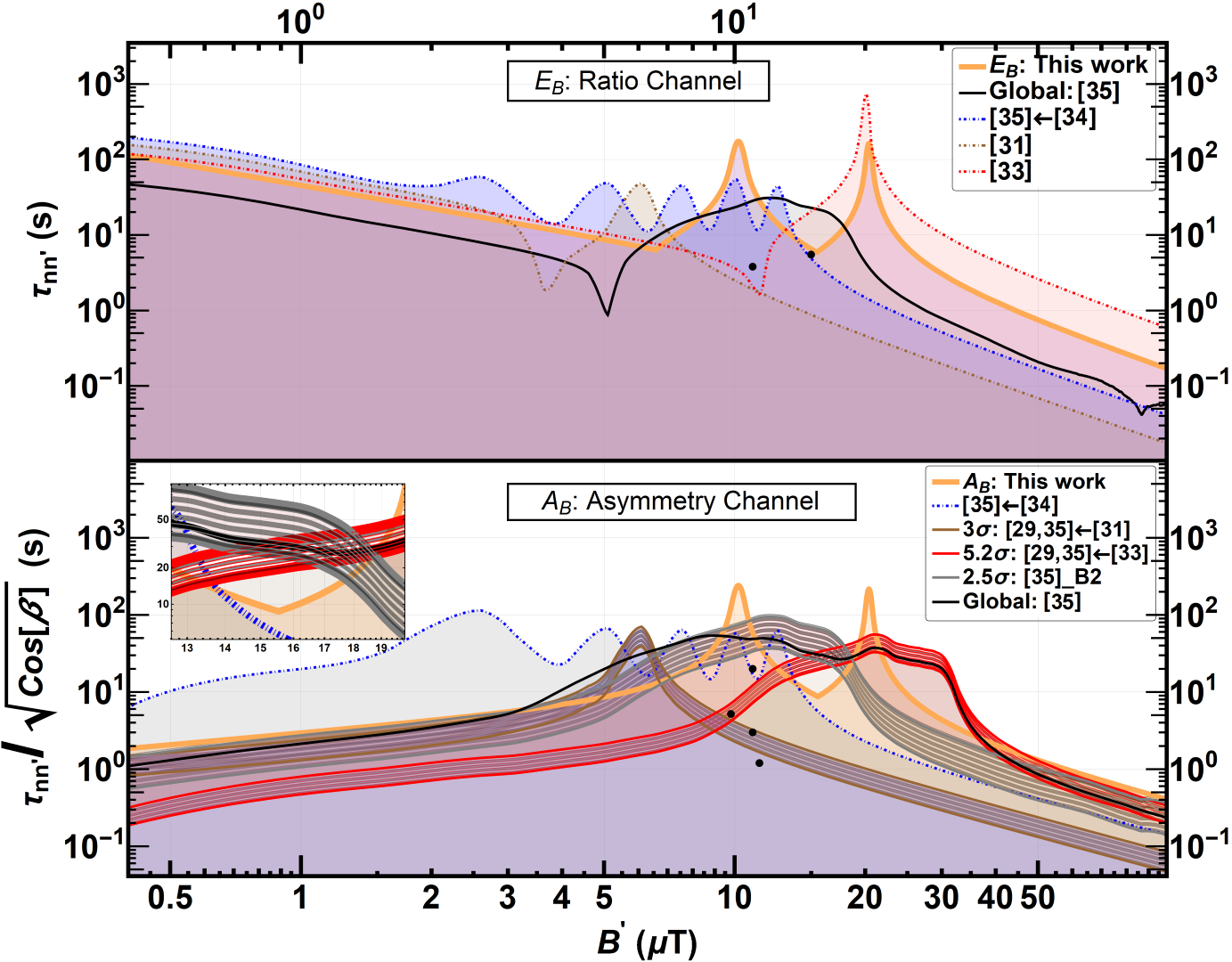} 
\caption[]{Lower limits on the $n-n'$ oscillation time, $\tau_{nn'}$ at 95\% C.L., using the ratio and asymmetry observables, while assuming $B'\ne0$. Top (bottom) panel shows the ratio (asymmetry) analysis, where the solid orange curve represents the lower limit on $\tau_{nn'}^{{B'}\ne0}$ ($\tau_{nn'}^{{B'}\ne0}/\sqrt{\cos\beta}$). (Top): The dot-dashed blue curve represents the lower limit imposed using data in Ref. \cite{[9-3]} by Ref. \cite{[8-4]}. The black curve represents the global constraint calculated by Ref. \cite{[8-4]} which imposes a weighted lower limit using data from Refs.~\cite{[9-1],[9-2-1],[9-3]} and the B2 series in Ref.~\cite{[8-4]}. The dot-dashed brown curve, represents the constraint from Ref. \cite{[9-1]}. The dot-dashed red curve represents the constraint from Ref. \cite{[9-2-1]}. The black dots indicate the solution consistent with the statistically significant signals as reported in Ref. \cite{[8-2]}. (Bottom): The black curve is the global constraint calculated in Ref. \cite{[8-4]}. The dot-dashed blue curve represents the lower limit imposed using data in Ref. \cite{[9-3]} by Ref. \cite{[8-4]}. The three striped regions are the signals (95\% C.L.): (i) the red striped region, is the signal region calculated in Refs. \cite{[8-3],[8-4]} from the $5.2\sigma$ anomaly in Refs. \cite{[9-2-1]}; (ii) the brown striped region is the signal calculated in Refs. \cite{[8-3],[8-4]} from the $3\sigma$ anomaly in Ref. \cite{[9-1]}; and (iii) the gray striped region is the signal from the $2.5\sigma$ anomaly observed in the B2 series of Ref. \cite{[8-4]}. The black dots indicate the solution consistent with the statistically significant signals as reported in Ref. \cite{[8-3]}. The inset shows an enlarged portion of the bottom plot between the ranges of $\SI{12.8}{\micro T}<B'<\SI{20}{\micro T}$.}
\label{fig4}
\vspace{-5mm}
\end{figure*}

The values of lower limits shown in Eqs.~\eqref{eq5-7-3-1-6}-\eqref{eq5-7-3-1-9} were scaled by $f_{E_B}(\eta)$ in Eq.~\eqref{eq4-2}, to generate a constraint plot in the parameter space defined by $\tau_{nn'}$ and $B'$. In this way two separate constraint curves were generated corresponding to $B\sim\{10,20\}\SI{}{\micro T}$. A lower envelop of the constraints obtained separately from the two curves is shown as our final constraint from the ratio analysis in Figure~\ref{fig4}~(Top).

\sloppy In the case of the asymmetry observable, Eq.~\eqref{eq5-2}, the function $f_{A_B}(\eta)$ does not change its sign. The lower limit of $\tau_{nn'}^{B'\ne0,A_B}/\left( \sqrt{f_{A_B}(\eta)}\cdot\sqrt{\cos\beta}\right)=1/\sqrt{-\left<D_B\right>}$ was obtained in a similar fashion to the above cases, from Eqs.~\eqref{eq6-3-1}-\eqref{eq6-3-2}, also at $95\%$ C.L.:
\begin{eqnarray}
\frac{\tau_{nn'}^{B'\ne0,A_B}}{\sqrt{\left|f_{A_B}(\eta)\right|}\cdot\sqrt{\cos\beta}} &> 4363~(B\sim\SI{10}{\micro T}) \label{eq11},\\
&> 3912~(B\sim\SI{20}{\micro T}). \label{eq12}
\end{eqnarray}

Our final constraint in the parameter space defined by $\left(\tau_{nn'}/\sqrt{\cos\beta}\right)$ and $B'$ from the asymmetry analysis is presented in Figure~\ref{fig4}~(Bottom) using the lower limits shown in Eqs.~\eqref{eq11}-\eqref{eq12} and scaling by $f_{A_B}(\eta)$. 

In Figure~\ref{fig4}, we also plotted the results from previous searches, including the signal-like anomalies listed in the caption. In case of a signal, in addition to the lower limit, a finite upper limit can be defined, making the confidence region a band along the $B'$ axis.

\section{Discussion}

The constraints from this work shown in Figure~\ref{fig4} (Top) and (Bottom) can be summarized as the following limits, respectively, at $95\%$ C.L.:
\begin{eqnarray}
\tau_{nn'}^{B'\ne0,E_B}&>&6~\text{s},~\SI{0.36}{\micro T}<B'<\SI{25.66}{\micro T}, \label{eq15}\\
\frac{\tau_{nn'}^{B'\ne0,A_B}}{\sqrt{\cos\beta}}&>&9~\text{s},~\SI{5.04}{\micro T}<B'<\SI{25.39}{\micro T}. \label{eq16}
\end{eqnarray}

The condition of $\omega'\left<t_f\right>^{(t_s)} \gg 1$, under which Eqs.~\eqref{eq4-2} and \eqref{eq5-2} are valid approximations, along with the value of $\left<t_f(t^*_s=180~\text{s})\right>=(0.0628\pm0.0027)~$s from Figure~\ref{fig1}, gives the lower bound of validity $B'>\SI{0.36}{\micro T}$ (at $95\%$ C.L.), on the horizontal axis of the plots in Figure~\ref{fig4}. The upper bound on the horizontal axis for the region of interest in Figure~\ref{fig4}, $B'<\SI{100}{\micro T}$, comes from constraints on UCN losses in the Earth's magnetic field~\cite{[8-2],[8-4]}.

According to Eqs.~\eqref{eq4-2} and \eqref{eq5-2}, the sensitivity to $n-n'$ oscillation has a singularity around $|B'-B|\sim0$, and was thus truncated in height according to Eq.~(6) in Ref.~\cite{[8-4]}. This behavior is responsible for the peaking of the solid curve in both plots in Figure~\ref{fig4} at $B'=\SI{10.20}{\micro T}$ and $B'=\SI{20.39}{\micro T}$. 

As in Ref.~\cite{[8-2]}, in this analysis, we considered that the mirror magnetic field $\bm{B'}$, and thus also $\beta$ are constant at the site of the experiment. While all the previous constraints on the $n-n'$ oscillation time come from experiments performed at the Institute Laue-Langevin (ILL) \cite{[8-2],[8-3],[9-1],[9-2],[9-2-1],[9-3],[8-4]} in Grenoble, France, our experiment was conducted at PSI in Villigen, Switzerland. A difference in $B'$ \emph{w.r.t.} the vertical between the geographic locations of PSI and ILL introduces an additional uncertainty when comparing exclusion plots from measurements at PSI and ILL, respectively. The comparison to results from ILL is valid under the natural assumption that a mirror magnetic field created within the Earth~\cite{[8-2]} displays approximate rotational symmetry, similar to the Earth's magnetic field. That is, its components change only on the level of 5\% between ILL and PSI~\cite{NOAA}, which would introduce a negligible offset on the horizontal axis of Figure~\ref{fig4}. In case the mirror magnetic field does not follow the Earth's rotation for various possible reasons, \emph{i.e.} due to a galactic mirror field, the observables would undergo a sideral modulation, an effect which was investigated in Ref.~\cite{[14]}.

In the ratio analysis, our constraint shown as a solid orange curve in Figure~\ref{fig4} (Top) is the best known constraint in the region $B'=\SI{10}{\micro T}$. In the asymmetry analysis, our constraint shown as a solid orange curve in Figure~\ref{fig4} (Bottom) excludes all signal spots (see black dots) reported in Ref. \cite{[8-3]}, for which our experiment was initially optimized.

It is important, however, to note that the three signal bands in the asymmetry analysis from Refs.~\cite{[8-3],[9-2-1],[8-4]} do not all overlap simultaneously, and thus exclude each other. Our analysis excludes three of the five regions where at least two of the signal bands overlap. Our result is also the best constraint at high mirror magnetic fields, $B'>\SI{37}{\micro T}$ in the asymmetry channel, along with being the best constraint around the mirror magnetic fields of $B'\sim\SI{10}{\micro T}$ and $B'\sim\SI{20}{\micro T}$. However, in the region of $\SI{4}{\micro T}<B'<\SI{37}{\micro T}$, our constraints do not exclude the signal bands of Ref.~\cite{[8-4]} which could be a focus of future efforts. The data for this experiment was collected in the summer of 2017. Even though our experiment was aimed at testing the signal-like anomalies indicated in Ref. \cite{[8-3]} (2012), it excludes significant portions of  the 2018 update of the signal-like anomalous regions in Ref. \cite{[8-4]}.

\section*{Acknowledgments}

We especially thank Z. Berezhiani for many valuable suggestions. The authors greatly acknowledge the exceptional support provided by Michael Meier, Fritz Burri and the BSQ group at PSI. The LPC and LPSC groups were supported by ANR grant \# ANR-14-CE33-0007-02. The University of Sussex group was supported by STFC grants \#ST/N504452/1, ST/M003426/1, and ST/N000307/1, and by their School of Mathematical and Physical Sciences. The PSI group was supported by SNSF grants \# 200020-137664, \# 200021-117696, \# 200020-144473, \# 200021-126562, \# 200020-163413 and \# 200021-157079. ETHZ was supported by SNSF grant \# 200020-172639. The University of Fribourg group was supported by SNSF grant \# 200020-140421. The University of Bern group was supported by the grants SNSF \# 181996 and ERC \# 715031-BEAM-EDM. The Jagiellonian University group was supported by the Polish National Science Center grant \# 2015/18/M/ST2/00056, \# 2016/23/D/ST2/00715 and \# 2018/30/M/ST2/00319. For the KU Leuven group, this work is also partly supported by Project GOA/2010/10 and  Fund for Scientific Research in Flanders (FWO). One of the authors, P. M., would like to acknowledge support from the SERI-FCS award \# 2015.0594 and Sigma Xi grants \# G2017100190747806 and \# G2019100190747806. We would like to acknowledge the grid computing resource provided by PL-GRID \cite{[plg]}. 


\end{document}